\def\be{\begin{equation}}
\def\ee{\end{equation}}
\def\ba{\begin{array}}
\def\ea{\end{array}}

\def\Rb{{I\!\! R}}

\def\Cb{\ \hbox{\vrule width 0.6pt height 6pt depth 0pt \hskip -3.5 pt} C}

\documentstyle[12pt]{article}
\begin{document}
\parskip=3pt
\parindent=18pt
\baselineskip=20pt
\setcounter{page}{1}
\centerline{\Large\bf A Remark on Symmetry of Stochastic Dynamical}
\vspace{2ex}
\centerline{\Large\bf Systems and Their Conserved Quantities}
\vspace{6ex}
\centerline{\large{\sf Sergio Albeverio$^\star$} ~~~and~~~ {\sf Shao-Ming Fei}
\footnote{\sf Alexander von Humboldt-Stiftung fellow.\\
\hspace{5mm}On leave from Institute of Physics, Chinese Academy of Sciences,
Beijing}}
\vspace{4ex}
\parindent=40pt
{\sf Institute of Mathematics, Ruhr-University Bochum,
D-44780 Bochum, Germany}\par
\parindent=35pt
{\sf $^\star$SFB 237 (Essen-Bochum-D\"usseldorf);
BiBoS (Bielefeld-Bochum);\par
\parindent=40pt
CERFIM Locarno (Switzerland)}\par
\vspace{6.5ex}
\parindent=18pt
\parskip=5pt
\begin{center}
\begin{minipage}{5in}
\vspace{3ex}
\centerline{\large Abstract}
\vspace{4ex}
Symmetry properties of stochastic dynamical systems described by stochastic
differential equation of Stratonovich type and related conserved
quantities are discussed, extending previous results by Misawa.
New conserved quantities are given by applying
symmetry operators to known conserved quantities.
Some detailed examples are presented.

\end{minipage}
\end{center}

\newpage
Symmetries and conserved quantities
have been discussed in the framework of Bismut's stochastic mechanics
\cite{bi} and Nelson's stochastic mechanics, see e.g. \cite{ne,gu,mo}.
More recently A.B. Cruzeiro, J.C. Zambrini, T. Kolsrud \cite{crk} and
Nagasawa \cite{na} have discussed stochastic variational principles
and associated conserved quantities in the theory of Schr\"odinger
processes(Euclidean quantum theory, in the sense of Zambrini, see also
\cite{az}). In \cite{misawa}(a stochastic version of \cite{hoj})
a theory of conserved quantities related to a stochastic
differential equation of Stratonovich type has been presented,
without referring to either Lagrangians or Hamiltonians.
In this letter we investigate the symmetry of the stochastic dynamical
differential equation and the space of conserved quantities. We derive
new results on conserved quantities which include the ones in
\cite{misawa}. It is shown that the conserved quantities are related to
the symmetry algebra of the space of conserved quantities.

We consider the stochastic dynamical systems of Stratonovich type\cite{ikeda}
described by the following n-dimensional vector valued stochastic
differential equations,
\be\label{1}
dx_t=b(x_t,t)dt+\sum_{r=1}^m g_r(x_t,t)\circ dw_t^r\,,
\ee
where $x_t$ is a $\Rb^n$-valued stochastic process,
$w_t=(w_t^r)_{r=1}^m$ is a $\Rb^m$-valued standard Wiener process,
$b=(b^i)_{i=1}^n$ and $g_r=(g_r^i)_{i=1}^n$ are $\Rb^n$-valued smooth
functions, $r=1,...,m$, satisfying restrictions
at infinity allowing the existence and uniqueness of solutions of (\ref{1}),
with given (deterministic or stochastic) initial condition $x_t\vert_{t=0}
=x_0$. Let ${\cal F}\equiv C^2(\Rb^n\times\Rb)$.
A function $I\in {\cal F}$ is called a conserved quantity of
stochastic dynamical system (\ref{1}) if it satisfies
\be\label{2}
\Delta_t I(x_t,t)=0\,,~~~~~\tilde{\Delta}_r I(x_t,t)=0\,,~~~~~~r=1,...,m,
\ee
where $\Delta_t =\partial_t +\sum_{i=1}^n b^i\partial_i$ and
$\tilde{\Delta}_r =\sum_{i=1}^n g_r^i\partial_i$, when $x_t$ satisfies
(\ref{1}). By Ito-Stratonovich's formula (\ref{2}) implies $d I(x_t,t)=0$ and
$I(x_t,t)=constant$ holds along the diffusion process $x_t$, the constant
being independent of $t$, but possibly depending on the
initial condition $x_0$. If
the initial condition $x_0$ in (\ref{1}) is taken to be deterministic,
then $I$ is a deterministic quantity independent of time.

In investigating the symmetry of the stochastic dynamical process (\ref{1}),
we would like to distinguish the symmetry of the stochastic differential
equation (\ref{1}) and the symmetry of the space of conserved quantities.
We first consider the former. Let $\epsilon > 0$
and $\zeta=(\zeta)_{i=1}^n\in {\cal F}$.

{\sf [Theorem 1]}. For $\epsilon$ sufficiently small the stochastic
differential equation (\ref{1}) is
invariant under following transformations
\be\label{3}
x_t^i\rightarrow x_t^i+\epsilon\zeta^i(x_t,t)\,,~~~~~i=1,...,n,
\ee
if $\zeta^i(x_t,t)$ satisfy
\be\label{4}
\ba{l}
\Delta_t\zeta^i(x_t,t)-\displaystyle\sum_{j=1}^n\zeta^j(x_t,t)
\partial_j b^i(x_t,t)=0\,,\\[3mm]
\tilde{\Delta}_r\zeta^i(x_t,t)-\displaystyle\sum_{j=1}^n\zeta^j(x_t,t)
\partial_j g_r^i(x_t,t)=0\,,~~~r=1,...,m.
\ea\ee

{\sf [Proof]}. Under (\ref{3}) equation (\ref{1})
becomes (writing shortly $\zeta$ for $\zeta(x_t,t)$):
$$\ba{rcl}
d(x_t^i+\epsilon\zeta^i)=dx_t^i+\epsilon d\zeta^i
&=&b^i(x_t+\epsilon\zeta,t)dt
+\displaystyle\sum_{r=1}^m g_r^i(x_t+\epsilon\zeta,t)\circ dw_t^r\\[4mm]
&=&b^i(x_t,t)dt+\displaystyle\sum_{j=1}^n\epsilon\zeta^j\partial_j b^i(x_t,t)
dt+
\displaystyle\sum_{r=1}^m g_r^i(x_t,t)\circ dw_t^r\\[4mm]
&&+\displaystyle\sum_{r=1}^m\sum_{j=1}^n
\epsilon\zeta^j\partial_j  g_r^i(x_t,t)\circ dw_t^r+o(\epsilon)\,,
\ea
$$
with $\epsilon^{-1}o(\epsilon)\to 0$ as $\epsilon\downarrow 0$. That is
$$
d\zeta^i=\sum_{j=1}^n\zeta^j\partial_j b^i(x_t,t) dt+
+\sum_{r=1}^m\sum_{j=1}^n\zeta^j\partial_j  g_r^i(x_t,t)\circ dw_t^r\,.
$$
On the other hand by the formula for Stratonovich differentials we have:
$$
d\zeta^i=\Delta_t\zeta^i dt+\sum_{r=1}^m\tilde{\Delta}_r\zeta^i\circ dw_t^r\,.
$$
Combining above two equations we obtain formulae (\ref{4}). $\rule{2mm}{2mm}$

Let $a^i, i=1,...,n$ and $a_0$ all belong to ${\cal F}$.
Then an operator $S=\sum_{i=1}^n a^i\partial_i +a_0\partial_t$ (acting on
$C^1(\Rb^n\times \Rb$)-functions)
is by definition a symmetry operator of the infinitesimal invariance
transformation (\ref{3}) of the stochastic equation (\ref{1}) if $S$
satisfies, on $C^1(\Rb^n\times \Rb$):
\be\label{5}
[S, x_t^i]=\zeta^i=a^i\,,
\ee
where $[\,,\,]$ is the Lie bracket and $\zeta^i$ satisfies equation (\ref{4}).

For the symmetry related to the space of conserved quantities of the
stochastic dynamical process (\ref{1}), we consider a linear operator $L$
satisfying the following commutation relations on $C^1(\Rb^n\times \Rb)$:
\be\label{6}
[\Delta_t,L]=T\Delta_t+\sum_{\alpha=1}^m T^\alpha\tilde{\Delta}_\alpha\,,~~~~~
[\tilde{\Delta}_r,L]=R_r\Delta_t+\sum_{\alpha=1}^m
R_r^\alpha\tilde{\Delta}_\alpha
\,,~~~r=1,...,m,
\ee
where $T,T^\alpha,R_r,R_r^\alpha\in {\cal F}$. Let
${\cal I}\equiv \{I(x_t,t)\vert dI(x_t,t)=0$ when $x_t$
satisfies (\ref{1})$\}$ be the space of the conserved functionals of the
process. We have

{\sf [Theorem 2]}. For $I\in {\cal I}$ and $L$ satisfying relation (\ref{6}),
$LI$ is also a conserved quantity, i.e., $LI\in{\cal I}$.

{\sf [Proof]}. As $I\in {\cal I}$, $I$ satisfies equation (\ref{2}).
{}From (\ref{6}) we further have $\Delta_t(LI)=0$,
$\tilde{\Delta}_r(LI)=0\,,~r=1,...,m$. Hence
$LI\in{\cal I}$. $\rule{2mm}{2mm}$

Let ${\cal L}$ denote the set of all operators $L$ satisfying (\ref{6}).

{\sf [Theorem 3]}. The set ${\cal L}$ is a complex Lie algebra under
Lie commutators(acting on $C^1(\Rb^n\times \Rb$)); that is, if
$L_1\,,L_2\,,L_3\in{\cal L}$, then
\parskip=0pt

(1) $a_1L_1+a_2L_2\in{\cal L}$, $\forall a_1,a_2\in\Cb\backslash\{0\}$,

(2) $[L_1,L_2]\in{\cal L}$,

(3) $[L_1,[L_2,L_3]]+[L_2,[L_3,L_1]]+[L_3,[L_1,L_2]]=0$.
\parskip=3pt

{\sf [Proof]}. Let $L_1\,,L_2\in{\cal L}$, s.t., on $C^1(\Rb^n\times \Rb$):
$$
[\Delta_t,L_i]=T_i\Delta_t+\sum_{\alpha=1}^m
T_i^\alpha\tilde{\Delta}_\alpha\,,~~~~~
[\tilde{\Delta}_r,L_i]=R_r^i\Delta_t
+\sum_{\alpha=1}^m R_r^{i\alpha}\tilde{\Delta}_\alpha\,,~~~~i=1,2\,.
$$
Property (1) is obviously as
$$
[\Delta_t,a_1L_1+a_2L_2]=(a_1T_1+a_2T_2)\Delta_t
+\sum_{\alpha=1}^m (a_1T_1^\alpha a_1+a_2T_2^\alpha)\tilde{\Delta}_\alpha
$$
and
$$
[\tilde{\Delta}_r,a_1L_1+a_2L_2]=(a_1R_r^1+a_2R_r^2)\Delta_t
+\sum_{\alpha=1}^m (a_1R_r^{1\alpha}+a_2R_r^{2\alpha})\tilde{\Delta}_\alpha\,.
$$
By a direct calculation we have, on $C^1(\Rb^n\times \Rb$):
$$
\ba{rcl}
[\Delta_t,[L_1,L_2]]&=&\left(L^1T_2-L^2T_1
+\displaystyle\sum_{\alpha=1}^m(T_1^\alpha R_\alpha^2
-T_2^\alpha R_\alpha^1)\right)\Delta_t\\[4mm]
&&+\displaystyle\sum_{\alpha=1}^m\left(L_1T_2^\alpha-L_2T_1^\alpha
+T_1T_2^\alpha-T_2T_1^\alpha
+\displaystyle\sum_{\beta=1}^m(T_1^\beta R_\beta^{2\alpha}
-T_2^\beta R_\beta^{1\alpha})\right)
\tilde{\Delta}_\alpha
\ea
$$
and
$$
\ba{rcl}
[\tilde{\Delta}_r,[L_1,L_2]]&=&\left(L^1R_r^2-L^2R_r^1
+\displaystyle\sum_{\alpha=1}^m(R_r^{1\alpha} R_\alpha^2
-R_r^{2\alpha}R_\alpha^1)\right)\Delta_t\\[4mm]
&&+\displaystyle\sum_{\alpha=1}^m\left(L_1R_r^{2\alpha}-L_2R_r^{1\alpha}
+R^1_rT_2^\alpha
-R_r^2T_1^\alpha
+\displaystyle\sum_{\beta=1}^m(R_r^{1\beta} R_\beta^{2\alpha}
-R_r^{2\beta}R_\beta^{1\alpha})\right)
\tilde{\Delta}_\alpha\,,
\ea
$$
where the linear property of operator $L_1$ and $L_2$ has been used.
Therefore we get $[L_1,L_2]\in {\cal L}$. The Jacobi identity (3)
is satisfied as $L\in{\cal L}$ are linear differential operators.
$\rule{2mm}{2mm}$

{}From Theorem 2 we have that the space of the conserved functionals admits
the symmetry algebra ${\cal L}$ in the sense that it is invariant
under any $L\in {\cal L}$. The space ${\cal I}$ is a representation
of the closed algebra ${\cal L}$. We call the elements of ${\cal L}$
``symmetry operators".

Now we consider a subalgebra ${\cal L}_0$ of ${\cal L}$ with
$T^r=R_r=0$, $R_r^\alpha=0$, for $\alpha\ne r$ and $R^r_r=T$ in relation
(\ref{6}). That is, for $L_0\in {\cal L}_0$, on $C^1(\Rb^n\times \Rb$):
\be\label{7}
[\Delta_t,L_0]=T\Delta_t\,,~~~[\tilde{\Delta}_r,L_0]=T\tilde{\Delta}_r\,,~~~r=1,...,m\,.
\ee
Let $L_0$ be of the form $L_0=\sum_{i=1}^n A^i\partial_i
+B\partial_t$ with $A^i, i=1,...,n$ and $B$ belong to ${\cal F}$.
A direct calculation
shows that relations (\ref{7}) are equivalent to the following equations:
\be\label{8}
\Delta_t B=T\,.
\ee
\be\label{9}
\Delta_t A^i-\sum_{j=1}^n A^j\partial_j b^i-B\partial_t b^i-Tb^i=0\,.
\ee
and
\be\label{10}
\tilde{\Delta}_r B=0\,,~~~r=1,...,m\,.
\ee
\be\label{11}
\tilde{\Delta}_r A^i-\sum_{j=1}^n A^j\partial_j g_r^i-B\partial_t g_r^i
-Tg_r^i=0\,,
\ee
as operators on $C^1(\Rb^n\times \Rb$).

We remark that in general a symmetry operator of the space ${\cal I}$
is not a symmetry operator of the infinitesimal transformations of the
stochastic differential equation. But
when $B=0$, then the equation set (8)-(11) reduces to the equation set
(\ref{4}) by replacing $A^i$ with $a^i$(=$\zeta^i$)
and $\sum_{i=1}^n A^i\partial_i$ is both a symmetry operator of the
infinitesimal invariance transformation of the stochastic
differential equation and for the space ${\cal I}$ of conserved quantities.

{\sf [Theorem 4]}. Let $L_0=\sum_{i=1}^n A^i\partial_i+B\partial_t \in
{\cal L}_0$. Then
\be\label{12}
I(x_t,t)=\sum_{i=1}^n \partial_i A^i(x_t,t)+\partial_t B(x_t,t)
-T(x_t,t)+L_0\phi(x_t,t)
\ee
is a conserved quantity of stochastic dynamical system (\ref{1})(i.e. for
$x_t$ satisfies (\ref{1})), when $\phi\in{\cal F}$ satisfies
\be\label{13}
\tilde{\Delta}_r\phi(x_t,t)+\sum_{i=1}^n \partial_i g_r^i(x_t,t)=0
\,,~~~r=1,...,m\,.
\ee
\be\label{14}
\Delta_t\phi(x_t,t)+\sum_{i=1}^n \partial_i b^i(x_t,t)=0\,.
\ee

{\sf [Proof]}. We have (dropping everywhere, for simplicity, the arguments
$x_t$, t):
$$
\ba{rcl}
\Delta_t I&=&
\Delta_t(\displaystyle\sum_{i=1}^n \partial_i A^i)+
\Delta_t(L_0\phi)+ \Delta_t(\partial_t B-T)\\[3mm]
&=&\displaystyle\sum_{i=1}^n \partial_i(\Delta_t A^i)-
\displaystyle\sum_{i,j=1}^n \partial_j b^i
\partial_i A^j + \Delta_t(L_0\phi)+ \Delta_t(\partial_t B-T)\\[3mm]
&=&\displaystyle\sum_{i,j=1}^n A^j\partial_j\partial_i b^i
+B\displaystyle\sum_{i=1}^n \partial_t\partial_i b^i
+\displaystyle\sum_{i=1}^n \partial_i B \partial_t b^i
+T\displaystyle\sum_{i=1}^n \partial_i b^i\\[3mm]
&&+\displaystyle\sum_{i=1}^n b^i\partial_i T
+T\Delta_t\phi+L_0\Delta_t\phi+\Delta_t(\partial_t B-T)\\[3mm]
&=&(L_0+T)(\displaystyle\sum_{i=1}^n \partial_i b^i+\Delta_t\phi)=0\,,
\ea
$$
$$
\ba{rcl}
\tilde{\Delta}_r I&=&
\tilde{\Delta}_r(\displaystyle\sum_{i=1}^n \partial_i A^i)+
\tilde{\Delta}_r(L_0\phi)+ \tilde{\Delta}_r(\partial_t B-T)\\[3mm]
&=&\displaystyle\sum_{i=1}^n \partial_i(\tilde{\Delta}_r A^i)-
\displaystyle\sum_{i,j=1}^n \partial_j g_r^i
\partial_i A^j + \tilde{\Delta}_r(L_0\phi)+ \tilde{\Delta}_r(\partial_t
B-T)\\[3mm]
&=&\displaystyle\sum_{i,j=1}^n A^j\partial_j\partial_i g_r^i
+B\displaystyle\sum_{i=1}^n \partial_t\partial_i g_r^i
+\displaystyle\sum_{i=1}^n \partial_i B \partial_t g_r^i
+T\displaystyle\sum_{i=1}^n \partial_i g_r^i \\[3mm]
&&+\displaystyle\sum_{i=1}^n g_r^i\partial_i T
+T\tilde{\Delta}_r\phi+L_0\tilde{\Delta}_r\phi+\tilde{\Delta}_r(\partial_t
B-T)\\[3mm]
&=&(L_0+T)(\displaystyle\sum_{i=1}^n \partial_i
g_r^i+\tilde{\Delta}_r\phi)=0\,,
\ea
$$
where eqs. (8-11), (13) and (14) have been used. Therefore by definition
$I(x_t,t)$ is a conserved quantity. $\rule{2mm}{2mm}$

Theorem 4 is a generalization of the one presented in \cite{misawa}, not only
because of the extra term $\partial_t B-T$, but also because of the presence
of $B$ in the symmetry operator $L_0$.
For the special case that $b^i=g^i_r$, or more generally $g^i_r=C(x_t,t)b^i$,
$r=1,...,m$, $\forall C(x_t,t)\in{\cal F}$ (these are the cases of the
examples given in \cite{misawa}),
we see from eqs. (\ref{8}) and (\ref{10}) that $\partial_t B-T=0$, hence
these terms disappear in the expression of $I(x_t,t)$.
Even in these cases formula (\ref{12}) is still a generalization of the one
in \cite{misawa} as long as $B\ne 0$ in $L_0$.

For a more detailed discussion we consider several examples.

$\underline{\bf Example~ 1}$ Following \cite{misawa} we consider the
3-dimensional stochastic linear dynamical system:
\be\label{15}
d\left(\ba{l}x_t^1\\[3mm]x_t^2\\[3mm]x_t^3\ea\right)=
\left(\ba{l}x_t^3-x_t^2\\[3mm]x_t^1-x_t^3\\[3mm]x_t^2-x_t^1\ea\right)dt+
\left(\ba{l}x_t^3-x_t^2\\[3mm]x_t^1-x_t^3\\[3mm]x_t^2-x_t^1\ea\right)\circ
dw_t\,.
\ee
In this case the existence and uniqueness of the solution is well know
(see e.g. \cite{ikeda}). The system (\ref{15}) has the properties
$g^i=b^i$, $\sum_{i=1}^3\partial_i g^i=0$ and
$\sum_{i=1}^3\partial_i b^i=0$. From eqs. (8-11) several solutions of
$L_0$ satisfying (\ref{7}) can be obtained with $T=0$. For instance
$$
\ba{l}
L_0=(x_t^1+x_t^2+x_t^3)\displaystyle\sum_{i=1}^3\partial_i\,,\\[3mm]
L_1=\left[(x_t^1)^2+(x_t^2)^2+(x_t^3)^2\right]
\displaystyle\sum_{i=1}^3\partial_i\,,\\[3mm]
L_2=(x_t^1x_t^2+x_t^2x_t^3+x_t^3x_t^1)\displaystyle\sum_{i=1}^3
\partial_i\,,\\[3mm]
L_3=\left[(x_t^1)^2(x_t^2+x_t^3)+(x_t^2)^2(x_t^1+x_t^3)
+(x_t^3)^2(x_t^2+x_t^1)
+3x_t^1x_t^2x_t^3\right]\displaystyle\sum_{i=1}^3\partial_i\,.
\ea
$$
By using Theorem 4 we can deduce that the following quantities are
conserved:
$$
\ba{l}
I_0=constant(independent~ of~ the~ x^i_t)\,,\\[3mm]
I_1=I_2=x_t^1+x_t^2+x_t^3\,,\\[3mm]
I_3=2((x_t^1)^2+(x_t^2)^2+(x_t^3)^2)+7(x_t^1x_t^2+x_t^2x_t^3+x_t^3x_t^1)\,,
\ea
$$
where $I_1$ is the conserved quantity obtained in \cite{misawa}.
$I_3$ is a new conserved quantity for the system (\cal{15}).

As $\Delta_t$ and $\tilde{\Delta}_r$ are linear operators, products of
conserved quantities are still conserved quantities. Let us set
$$
I_3^\prime\equiv(I_3-2I_2^2)/3=x_t^1x_t^2+x_t^2x_t^3+x_t^3x_t^1\,.
$$
$I_1$ and $I_3^\prime$ are then two simple nontrivial conserved quantities of
the system (\cal{15}). Symmetry operators map conserved quantities
into conserved quantities. Under the actions of the symmetry operators
$L_i, i=1,2,...$, we have, e.g.,
$$
\ba{l}
L_0 I_1=3I_1\,,~~~L_1 I_1=3(I_1^2-2I_3^\prime)\,,\\[3mm]
L_2 I_1=3I_3^\prime\,,~~~L_0 I_3^\prime=2I_1^2\,.
\ea
$$

In fact $L_0=I_1\sum_{i=1}^3\partial_i$ and
$L_2=I_3^\prime\sum_{i=1}^3\partial_i$.  In the present case $r=1$ and
$$
\left[\sum_{i=1}^3\partial_i,\Delta_t\right]=0\,,~~~
\left[\sum_{i=1}^3\partial_i,\tilde{\Delta}_1\right]=0\,,
$$
on $C^1(\Rb^3\times \Rb$) functions.
Let $f$ be an arbitrary polynomial function on $\Rb^2$.
Since $L\equiv f(I_1,I_3^\prime)\sum_{i=1}^3\partial_i$ commutes with
$\Delta_t$ and $\tilde{\Delta}_1$, we have that $L$
is a symmetry operator in ${\cal L}_0$(defined in (\ref{7})).
Hence the system (\ref{15}) possesses an infinite number of symmetry
operators that are linear independent. They constitute an algebra
${\cal L}_0$ with commutation relations which can obviously be explicitly
computed, e.g.,
$$
[L_0,L_1]=4L_3-L_1\,,~~~[L_0,L_2]=2L_1+L_3
$$
on $C^1(\Rb^n\times \Rb$).
As $B=0$ in this example, the algebra ${\cal L}_0$ coincides with the
algebra generating the infinitesimal invariance transformations
for the stochastic differential equation.

The following examples are dedicated to show the useful symmetry analysis
on conserved functionals in stochastic dynamical systems.
We start with a (unique) solution for small times and show that associated
to it there are conserved quantities (functionals of the solution process).

$\underline{\bf Example~ 2}$ Let us consider the following stochastic
dynamical system
\be\label{15.5}
dx_t^i=b^i(x_t,t)dt+ g^i(x_t,t)\circ dw_t
\ee
with
$$
g^i=x_t^i((x_t^1)^2+(x_t^2)^2+(x_t^3)^2)^m(e^{2mt}\delta_{n,0}
-\frac{1}{nt}\delta_{n,-2m})\,,
$$
$$
b^i=\frac{x_t^i}{(x_t^1+x_t^2+x_t^3)^n}(e^{2mt}\delta_{n,0}-b_0\delta_{n,-2m})\,,
{}~~~i=1,2,3,
$$
where $n\,,m\in Z$, $m\ne 0$, and $b_0\in \Rb$. We have the symmetry operators
in ${\cal L}_0$ satisfying (on $C^1(\Rb^n\times \Rb$))
$$
[\Delta_t,L_i]=T_i\Delta_t\,,~~~[\tilde{\Delta}_1,L_i]=T_i\tilde{\Delta}_1\,,
{}~~~i=1,2,3,
$$
with
\be\label{16}
\ba{l}
L_1=x_t^3\partial_2-x_t^2\partial_3
+\displaystyle\frac{x_t^2-x_t^3}{x_t^1+x_t^2+x_t^3}
(e^{-2mt}\delta_{n,0}-nt\delta_{n,-2m})\partial_t\,,
\\[4mm]
L_2=x_t^1\partial_3-x_t^3\partial_1
+\displaystyle\frac{x_t^3-x_t^1}{x_t^1+x_t^2+x_t^3}
(e^{-2mt}\delta_{n,0}-nt\delta_{n,-2m})\partial_t\,,
\\[4mm]
L_3=x_t^2\partial_2-x_t^1\partial_2
+\displaystyle\frac{x_t^1-x_t^2}{x_t^1+x_t^2+x_t^3}
(e^{-2mt}\delta_{n,0}-nt\delta_{n,-2m})\partial_t
\ea
\ee
and
$$
\ba{l}
T_1=\displaystyle\frac{2m(x_t^3-x_t^2)}{x_t^1+x_t^2+x_t^3}
(e^{-2mt}\delta_{n,0}-\delta_{n,-2m})\,,\\[4mm]
T_2=\displaystyle\frac{2m(x_t^1-x_t^3)}{x_t^1+x_t^2+x_t^3}
(e^{-2mt}\delta_{n,0}-\delta_{n,-2m})\,,\\[4mm]
T_3=\displaystyle\frac{2m(x_t^2-x_t^1)}{x_t^1+x_t^2+x_t^3}
(e^{-2mt}\delta_{n,0}-\delta_{n,-2m})\,,
\ea
$$
where $\Delta_t=\sum_{i=1}^3 b^i\partial_i +\partial_t$ and
$\tilde{\Delta}_1=\sum_{i=1}^3 g^i\partial_i$.

A function $\phi$ satisfying (\ref{13}) is given by
$$
\phi(x_t,t)=-\frac{3+2m}{3}\log{(x_t^1x_t^2x_t^3)}+e^{2mt}\delta_{n,0}\,,
$$
for $x_t^i\neq 0,~i=1,2,3$. From Theorem 4 we have the conserved quantities:
$$
\ba{l}
I_1 =\displaystyle \frac{(2m+3)((x_t^2)^2-(x_t^3)^2)}{3x_t^2x_t^3}

+\displaystyle\frac{2m(x_t^2-x_t^3)}{x_t^1+x_t^2+x_t^3}\delta_{n,0}\,,\\[4mm]
I_2 = \displaystyle\frac{(2m+3)((x_t^3)^2-(x_t^1)^2)}{3x_t^1x_t^3}

+\displaystyle\frac{2m(x_t^3-x_t^1)}{x_t^1+x_t^2+x_t^3}\delta_{n,0}\,,\\[4mm]
I_3 = \displaystyle\frac{(2m+3)((x_t^1)^2-(x_t^2)^2)}{3x_t^1x_t^2}
	 +\displaystyle\frac{2m(x_t^1-x_t^2)}{x_t^1+x_t^2+x_t^3}\delta_{n,0}\,.
\ea
$$
In this example the terms $\partial_tB_i-T_i,~i=1,2,3$ appearing in
(\ref{12}) are zero. But
as $B_i\ne 0$, the term $L_i\phi$ still contributes extra terms to $I_i$.

The space of conserved quantities of the system (\ref{15.5}) is $SU(2)$
symmetric. It is straightforward to check that
the symmetry operators (\ref{16}) satisfy the $SU(2)$ algebraic relations:
$$
[L_i,\,L_j]=\epsilon_{ijk}L_k\,,~~~~i,j,k=1,2,3.
$$

$\underline{\bf Example~ 3}$
The following example is a nonlinear model with $\partial_tB-T\ne 0$
in (\ref{12}).
\be\label{17}
d\left(\ba{l}x_t^1\\[3mm]x_t^2\\[3mm]x_t^3\ea\right)=
\frac{1}{t}\left(\ba{l}x_t^1\\[3mm]x_t^2\\[3mm]x_t^3\ea\right)dt+
\frac{1}{t}\left(
\ba{l}x_t^1(x_t^3-x_t^2)\\[3mm]x_t^2(x_t^1-x_t^3)\\[3mm]
x_t^3(x_t^2-x_t^1)\ea\right)\circ dw_t\,,~~t>0\,.
\ee
For this system we have a symmetry operator $L_0\in {\cal L}_0$ given by
$$
L_0=B\partial_t=(x_t^1+x_t^2+x_t^3)\partial_t
$$
satisfying
$$
\ba{l}
[\Delta_t,L_0]=T\Delta_t=
\displaystyle\frac{1}{t}(x_t^1+x_t^2+x_t^3)\Delta_t\,,\\[4mm]
[\tilde{\Delta}_1,L_0]=T\tilde{\Delta}_1=
\displaystyle\frac{1}{t}(x_t^1+x_t^2+x_t^3)\tilde{\Delta}_1\,,
\ea
$$
where
\be\label{19}
\Delta_t=\partial_t+\sum_{i=1}^3 \frac{x_t^i}{t}\partial_i\,,~~~
\tilde{\Delta}_1=\frac{x_t^1(x_t^3-x_t^2)}{t}\partial_1
+\frac{x_t^2(x_t^1-x_t^3)}{t}\partial_2
+\frac{x_t^3(x_t^2-x_t^1)}{t}\partial_3\,.
\ee
$\phi$ satisfying (\ref{13}) and (\ref{14}) is given by $\phi=-3\log t,
{}~t\ne 0$. From Theorem 4 we have, for $x_t$ satisfying (\ref{17}):
$$
I(x_t,t)=\sum_{i=1}^n \partial_i A^i+\partial_t B-T+L_0\phi
=-\frac{4}{t}(x_t^1+x_t^2+x_t^3)\,,~~~~t > 0\,.
$$

Let us summarize.
By investigating the symmetry of the space of conserved quantities for
stochastic dynamical systems, we have established new relations for
conserved quantities. We would like to indicate that although the conserved
functionals are given by the elements of a subalgebra ${\cal L}_0$
of ${\cal L}$, the space of conserved functionals itself admits
the symmetry Lie algebra ${\cal L}$. Let us consider (\ref{17}),
with $\Delta_t$ and $\tilde{\Delta}_1$ given by (\ref{19}), as an example.
We consider the operator
$L=a(x_t,t)\Delta_t+b(x_t,t)\tilde{\Delta}_1$, $a(x_t,t),b(x_t,t)\in{\cal F}$.
Noting that in present case $[\Delta_t,\tilde{\Delta}_1]=0$, we have
$$
[\Delta_t,L]=\Delta_t a(x_t,t)\Delta_t+\Delta_t b(x_t,t)\tilde{\Delta}_1\,,~~~
[\tilde{\Delta}_1,L]=\tilde{\Delta}_1 a(x_t,t)\Delta_t
+\tilde{\Delta}_1 b(x_t,t)\tilde{\Delta}_1\,.
$$
Therefore $L$ is a symmetry operator in ${\cal L}$ which maps $I\in{\cal I}$
to zero. However only when $\Delta_t b(x_t,t)=\tilde{\Delta}_1 a(x_t,t)=0$
and $\tilde{\Delta}_1 b(x_t,t)=\Delta_t a(x_t,t)$ is $L$ a symmetry
operator in ${\cal L}_0$, satisfying the defining relations (\ref{7}).

\vspace{4ex}
ACKNOWLEDGEMENTS: We would like to thank the A.v. Humboldt
Foundation for the financial support given to the second named author.

\end{document}